# Evolution of static to dynamic mechanical behavior in topological nonreciprocal robotic metamaterials

Zehuan Tang [a], Tingfeng Ma [b], Hui Chen [b], Yuanwen Gao [a,*]

[a] Key Laboratory of Mechanics on Environment and Disaster in Western China, The Ministry of Education of China, Lanzhou University, Lanzhou 730000, China

[b] School of Mechanical Engineering and Mechanics, Ningbo University, Ningbo 315211, China

## ABSTRACT

Based on the Maxwell-Beatty reciprocity theorem, static non-reciprocity has been realized by using nonlinearity, but this non-reciprocity has strict restrictions on input amplitude and structure size (number of units). Here, we propose a robotic metamaterial with two components of displacement and rotation, which uses active control to add external forces on the units to break reciprocity at the level of the interactions between the units. We show analytically and simulatively that breaking reciprocity at the level of the interactions directly leads to a strong asymmetric response of displacement in a static system, this displacement-specific characteristic not only has no restrictions on size, input amplitude, and suitable geometric asymmetry, but also can be transmitted to rotation by coupling under large deformation. After the evolution from statics to dynamics, asymmetric transmission and unidirectional amplification of vector solitons are both implemented in this system. Our research uncovers the evolution of static non-reciprocity to dynamic non-reciprocity while building a bridge between non-reciprocity physics and soliton science.

# 1. Introduction

Reciprocity is a fundamental property of linear, time-reversal invariant, and spatial-reversal invariant physical systems, in which the excitation and response are symmetric[1]. Reciprocity can also sometimes be seen as a hindrance, for example in the case of reciprocity, there is no way to tune the response to different levels under excitation from different sides. Breaking this limitation of reciprocity is expected to achieve full control over static and dynamic responses. In the last few years, non-reciprocity has been achieved in various mechanical systems[2-6], which achieve non-reciprocity primarily through two schemes. One is to break the time-reversal symmetry of the system by adding active time-modulated components, which are generally used to achieve dynamic non-reciprocity. Typical examples are active metamaterials[7-10] and gyroscopic metamaterials[11, 12], which realize the propagation of non-reciprocal waves. The other is to break spatial symmetries of the passive structure through nonlinearity, which extends non-reciprocity to statics. For example, in topologically trivial "fishbone" metamaterials[13-15], the structure has an asymmetrical response under different side excitation, while in topologically non-trivial rotating square metamaterials[13, 16], the response has greater asymmetry under different side excitation. This not only proves that breaking spatial symmetry is a powerful means to achieve non-reciprocity, but also shows that structures with topological properties can achieve more significant non-reciprocity.

In conventional topological band theory, the principle of bulk-edge correspondence (BEC) directly relates the number of topological boundary modes to the topological invariants under periodic boundary conditions (PBC)[17, 18]. The combination of non-reciprocity and topological band theory extends conventional band topology, such as zero modes based on the geometric asymmetry[13] and gain/loss induced nontrivial topological phases[19-21], and reveals a new class of topological phases characterized by non-Bloch winding numbers[22-24]. Perhaps the phenomenon of most interest in this novel spectral topology is the non-Hermitic skin effect (NHSE), where eigenstates in a system with broken reciprocity decay exponentially from the system boundary, and all eigenstates are localized at the boundary of the finite-sized system[25]. In the field of metamaterials, NHSE is generally achieved by adding active modulated components, which inevitably inject/extract energy into a closed system.

Recently, the idea of combining active metamaterials and robotics has been proposed, and robotic metamaterials that combine local sensing, computation, communication, and actuation functions have been designed[26], which can achieve more extreme performance and different combinations of performance[27]. The concept of robotic metamaterials stems from the "programmability problem"[28], whereby shape-changing robots designed in this way are often complex because the microprocessors inside each robot need to communicate with all the other microprocessors in the lattice. Since then, the design idea of the common command instruction has greatly simplified the design of robot metamaterials, when the material is loaded externally, each robot only needs to respond to the information transmitted to it by the neighboring units according to the common instruction[29]. Based on this, we design a robotic mechanical metamaterial that uses a distributed active control to break reciprocity at the level of the interactions between units, which results in the spatial symmetry of the structure being broken. This robotic metamaterial compares to conventional active or passive metamaterials, in addition to having a unique function that allows standing waves to achieve unidirectional amplification at all frequencies[26], our study shows that it also has a range of more extreme properties in static behavior, such as the non-reciprocity is not limited by structural size, the magnitude of input amplitude and geometric asymmetry. In particular, in the current study of vector soliton regulation, both active and passive metamaterials basically change the bond parameters, either directly or indirectly[30, 31]. Therefore, besides using the amplitude gap of vector solitons to realize the unidirectional transmission[32], usually unable to achieve asymmetric transmission under excitation from different sides. However, the robotic metamaterial we designed is based on the perspective of energy injection/extraction into the system, it can realize both unidirectional amplification and asymmetric transmission of vector solitons.

## 2. Non-reciprocal robotic metamaterials

The one-dimensional chain consists of rigid cross units with mass *m*, rotational inertia *J*, and arm length *l*, and the crosses are connected by shims of length $l_h$. Each cross is equipped with a local control device, which applies a force related to the displacement of neighboring units on the cross, the non-reciprocity at the level of the interactions can be achieved when the force and displacement satisfy the appropriate relationship. The specific design scheme is as follows: the displacement sensors on the (*n*-1)-th cross and (*n*+1)-th cross needs to collect their displacement signals $u_{n\pm1}$, then the collected displacement signals are transmitted via the wire to the microprocessor on the *n*-th cross, and the motor on the *n*-th cross finally applies a rightward force $F_n = k_n(u_{n-1} - u_{n+1})$ to the *n*-th cross, as shown in Fig. 1(a). For mathematical modeling of the system, $F_n$ is modeled as the resultant force of the leftward force $F_{n,n-1}$ and the rightward force $F_{n,n+1}$, as shown in Fig. 1(b), which satisfies the resultant relation $F_{n,n+1} - F_{n,n-1} = F_n$. It is worth mentioning that in order to explain the problem more easily in different cases, $F_{n,n-1}$ and $F_{n,n+1}$ can be modeled in any form if the condition of the resultant relation is satisfied. For example, under the PBC, $F_{n,n-1} = k_n(u_n - u_{n-1})$ and $F_{n,n+1} = -k_n(u_{n+1} - u_n)$ are selected for the modeling of the equivalent system, a pair of actuation forces between neighboring units in Fig. 1(b) can be equivalent to the interaction force generated by the spring connection. The spring will have a special property —— the force at both ends of the spring is not equal —— namely, when the *n*-th node acts on the (*n*+1)-th node, the stiffness is $k_n$, while when the (*n*+1)-th node acts on the *n*-th node, the opposite negative stiffness $-k_n$ is presented[26], the modeling of the equivalent system is shown in Fig. 1(c). In this equivalent system, reciprocity at the level of the interactions is violated, which provides conditions for the realization of non-reciprocal effects. In addition, as shown in Fig. 1(c), shims connecting the crosses are modeled as a combination of tensional, torsional, and shearing springs with stiffness $k_l$, $k_\theta$, and $k_s$ [33], and the tension stiffness $k_l$ and shear stiffness $k_s$ are much greater than the torsion stiffness $k_\theta$.

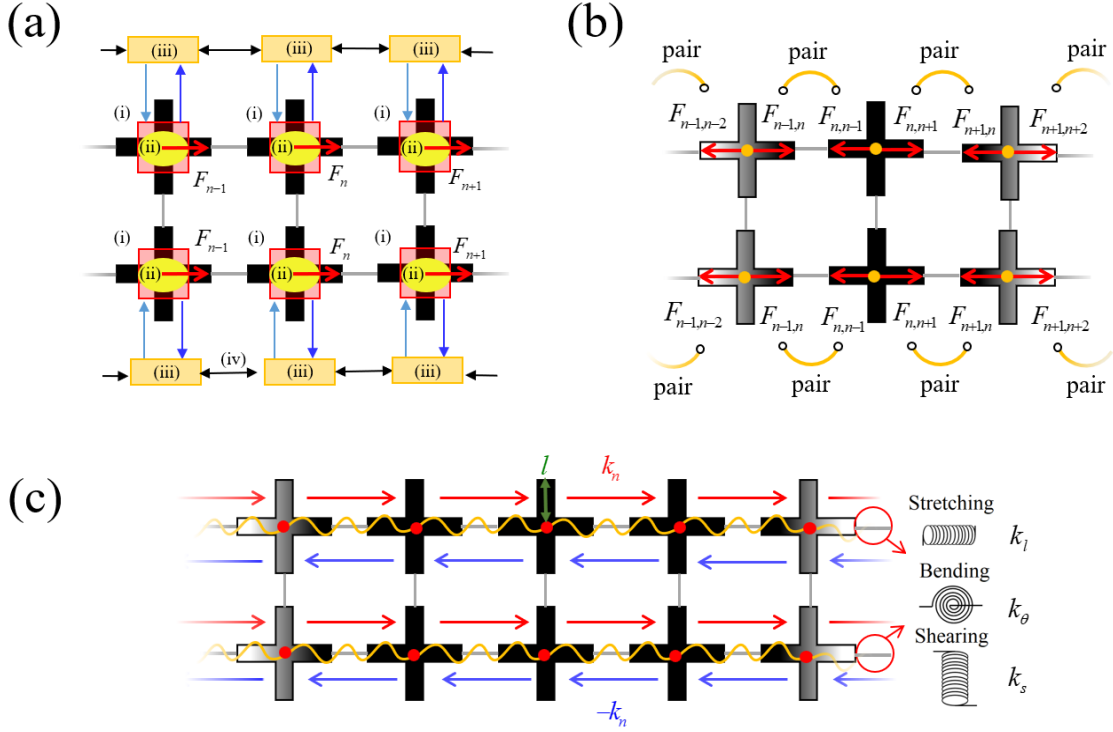

Fig. 1. (a) Design drawing of nonreciprocal robotic metamaterials, each unit is a minimal robot with two degrees of freedom (translational and rotational), which includes a displacement sensor (i), a DC motor (ii), and a microprocessor (iii). Each unit transmits signals to its neighboring units via wires (iv). (b) Modeling diagram of the system. (c) Modeling diagram of the equivalent system under PBC.

### 3. Static non-reciprocity

3.1. Pure displacement framework

First, the case where the actuation forces of both the left and right ends are at the center of mass is discussed, namely, the structure is in the pure displacement situation without rotation angle. The structure can be regarded as a one-dimensional mass-spring system, and the governing equation of displacement is as follows:

$$m\frac{\partial^2 u_n}{\partial t^2} = k_n(u_{n-1} - u_{n+1}) + k_l(u_{n+1} + u_{n-1} - 2u_n). \tag{1}$$

In general, the band theory is used to analyze the system under PBC, and the winding number of a closed loop formed by branches can effectively represent the number of boundary modes of the finite structure[34, 35]. However, due to the possible skin effect in non-reciprocity systems, the traditional principle of BEC may fail in systems where reciprocity is broken[25, 36]. In our non-reciprocal system, the winding number under PBC produces singular results that cannot be used to characterize topological properties (see Appendix A for details). The specific topological properties of non-reciprocal systems will need to be conducted under open boundary conditions (OBC). The modeling diagram of the finite structure of the $N'(N' = N+1)$ nodes is shown in Fig. 2(a). Under the condition that the overall horizontal external force balance is satisfied, namely:

$$F_{0,1} + \sum_{1}^{N-1}(F_{n,n+1} - F_{n,n-1}) - F_{N-1,N} = 0,$$

the general formulas of actuation force can be written as $F_{n,n-1} = -k_n(u_n + u_{n-1})$ and $F_{n,n+1} = -k_n(u_{n+1} + u_n)$, so the governing equation of the finite structure is:

$$-m\omega^2 U = D_{N' \times N'} U, \qquad (2)$$

where, $D_{N' \times N'}$ is the total stiffness matrix of the finite structure (see Appendix A for specific form). The result of solving the 30 nodes is shown in Fig. 2(b), an isolated band appears at the zero frequency in the band gap, as shown by the red solid line in the figure. For the analysis of this isolated band, Eq. (2) degenerates to $D_{N' \times N'} U=0$ at zero frequency, and $\{u_n\}$ is a geometric series, namely:

$$u_n = u_0[g(k_n)]^n, \qquad (3)$$

where, $g(k_n) = c_1/c_2 = (k_l + k_n)/(k_l - k_n)$ is the local factor of the mode.

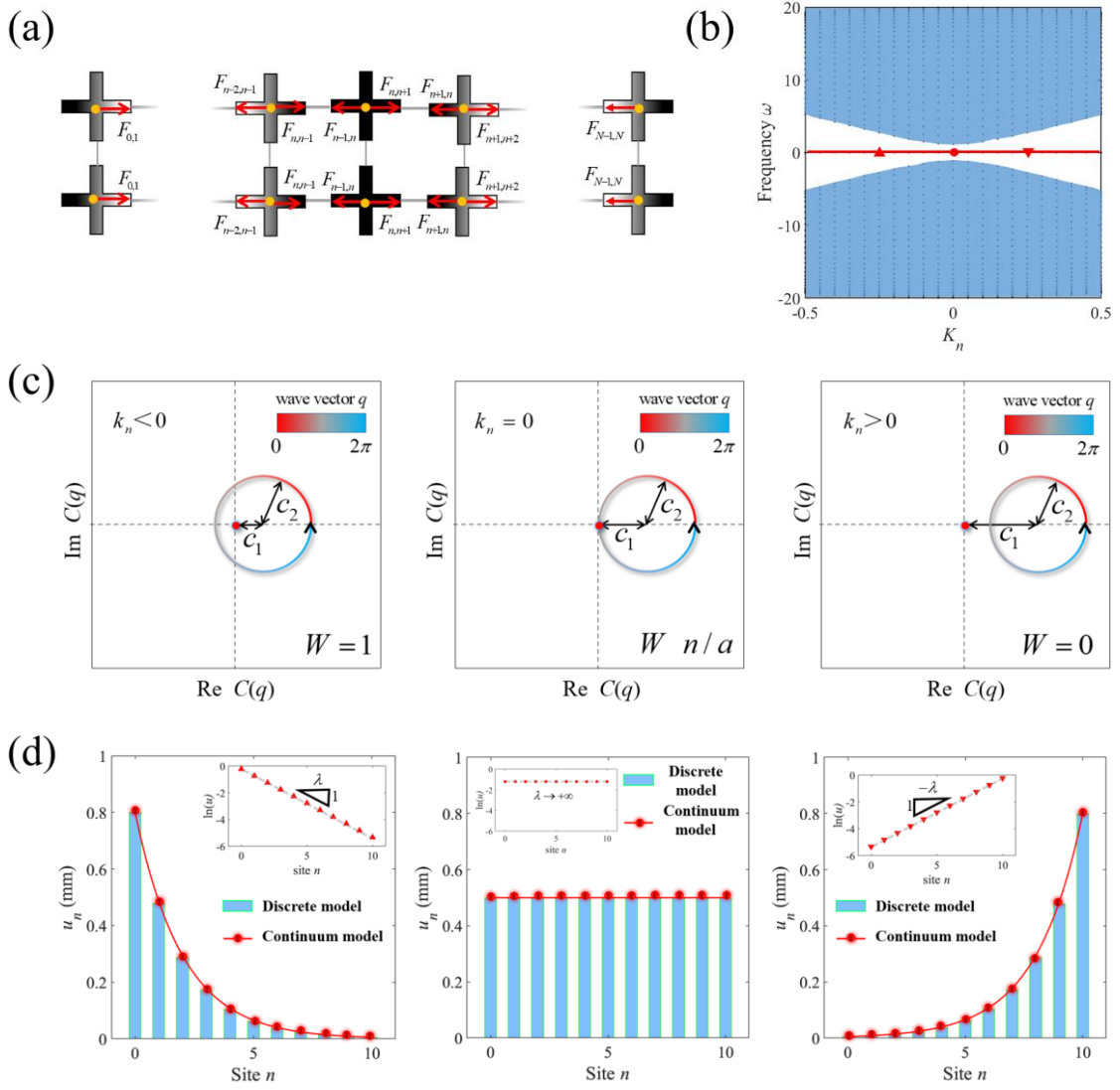

Fig. 2. (a) Schematic diagram of the OBC. (b) Spectra of the structures under the OBC. (c) The topological invariant changes with $k_n$. (d) The distribution of the modes in the finite structure with 11 nodes, inset is the distribution diagram after linearization.

The above phenomena are further analyzed, in the deformation process, the work done by the $n$-th pair of actuation forces ($F_{n,n+1}$ and $F_{n+1,n}$) is:

$$E_{n,n+1} = \int F_{n,n+1} du_n - \int F_{n+1,n} du_{n+1} = k_l(g-1)^2 u_n^2 / 2,$$

the strain energy required for the shim deformation between the $n$-th and $(n+1)$-th node is:

$$U_{n,n+1} = k_l(u_{n+1} - u_n)^2 / 2 = k_l(g-1)^2 u_n^2 / 2,$$

namely, $E_{n,n+1} - U_{n,n+1} = 0$. This shows that the work injected by the actuation force between units to the chain is exactly all used for the deformation of the shim between corresponding units. If the control device and the one-dimensional chain are equivalent to a system, the system supports a mode similar to the zero-energy modes, this corresponds well to the previously reported topological zero-energy modes induced geometric asymmetry[37, 38]. According to the above energy equality relationship, $F_{n,n+1}$ and $F_{n+1,n}$ should be equal to the reaction force generated by the tensioned shim, namely, $F_{n,n+1} = F_{n+1,n} = -k_l(u_{n+1} - u_n)$, for a finite structure with $N'$ nodes, write the above formula in matrix form:

$$C_{(N'-1) \times N'} U = 0, \tag{4}$$

where, $C_{(N'-1) \times N'}$ is the compatibility matrix (see Appendix A for the specific form). The Fourier transform of the compatibility matrix C yields $C(q) = c_1 - c_2 e^{iq}$, when the wave vector $q$ changes from 0 to $2\pi$, $C(q)$ forms a circle in the complex plane. The circle has an obvious spectral winding number, which is defined as $W = (2\pi i)^{-1} \int_0^{2\pi} 1/C(q) dC(q)$. Fig. 2(c) shows the change of topological invariants with $k_n$. When $k_n < 0$ ($g < 1$), the circle goes once around the origin, and the topological invariant $W=1$, in the case of $N \to +\infty$, $u_n$ is an admissible eigenstate near the boundary $n=N$; When $k_n = 0$ ($g = 1$), the circle passes through the origin, at this point, $W$ is singular, with a topological transition in the one-dimensional chain. When $k_n > 0$ ($g > 1$), the topological invariant $W$ changes to 0, $u_n$ is not an admissible eigenstate near the boundary $n=N$. Fig. 2(d) shows the distribution of displacements corresponding to zero frequency for a finite structure with 11 nodes, the insets are the linearized displacement distribution diagram. It can be seen that when $k_n \neq 0$, the modes are exponentially distributed, and the decay length $\lambda$ is defined as the length required to decay a unit in the logarithmic coordinate, namely, $\lambda = 1/\ln(u_n/u_{n+1})$, and the decay length reflects the decay speed of the boundary mode. When $W=1$, the decay length is positive, and the mode is localized on the left. When $W$ is singular, the decay length tends to positive infinity, and the modes are delocalized, which appear as body modes. When $W=0$, the decay length is negative, displacement increases exponentially with the increase of the node numbering, and the mode is localized on the right. With the $k_n$ symbol unchanged, the modes only localize on a specific side to provide conditions for the implementation of static non-reciprocity.

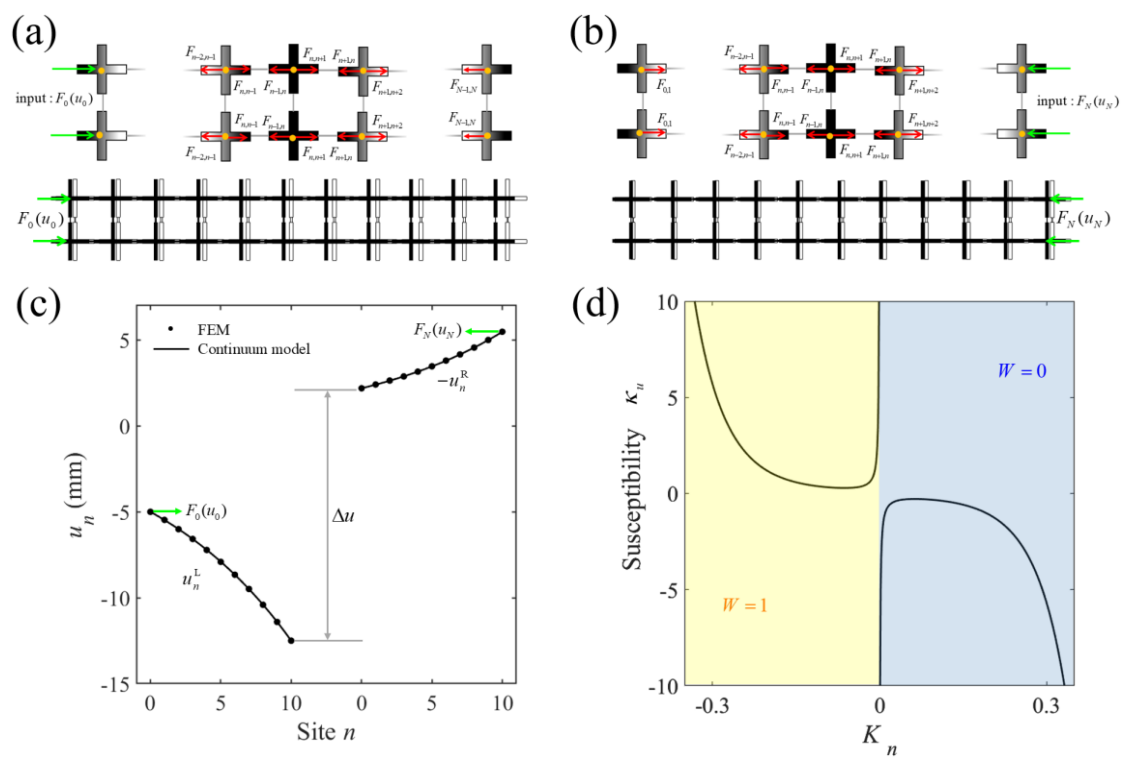

Fig. 3. (a) and (b) The results of the left and right excitation at $k_n/k_l = 0.05$, respectively. The top figures are the excitation schematic diagram. the lower figures are the results of the finite element model (FEM), the crosses not filled with the color are the initial state, and the crosses filled by the color are the deformation after the excitation. When the left side is excited, the displacement increases from the excitation side to the right side. When the right side is excited, the displacement tends to decay from the excitation side to the left side. (c) The static non-reciprocity of the displacement. (d) The non-reciprocal susceptibility $\kappa_u$ diverges at $K_n = 0$, where $K_n = k_n/k_l$.

A chain with $N'$ nodes is created in the simulation software COMSOL, without general we choose $N' = 11$, and the built-in probe function is used to monitor the centroid displacement $u_n$, monitored displacement $u_n$ to add actuation force on each node, the way of actuation force described in Fig. 2(a), but the difference is the start nod (the first node for the left incentive and the last node for the right excitation) without actuation force. First, the one-dimensional chain is excited on the left, and a rightward force $F_0(u_0)$ is input at the first node to deform the chain. The results of the finite element model (FEM) show that the displacements of each node tend to increase with the increase of node numbering, as shown in Fig. 3(a). A leftward force $F_N(u_N)$ is input at the last node, and the displacement decays from the excitation side to the left side, as shown in Fig. 3(b). In the case of $F_0 = F_N$, there is a huge difference between $u_n^L$ under left excitation and $-u_n^R$ under right excitation, which indicates that the displacement response has obvious static non-reciprocity, as shown in Fig. 3(c). The displacement difference between left and right excitation increases with the increase of excitation amplitude and structure size, and the relationship between them is $\Delta u = \kappa_u(k_n)F_0$ (see Appendix A for details), $\Delta u$ and $F_0$ present a primary dependence relationship, where $\kappa_u$ is non-reciprocal susceptibility, and its specifically expressed is as follows:

$$\kappa_u(k_n) = -\frac{g^N + g^{1-N}}{k_l(g-1)}. \tag{5}$$

The variation of $\kappa_u$ concerning to $k_n$ is shown in Fig. 3(d), it can be seen that when $k_n$ changes from negative to positive, $\kappa_u(k_n)$ appears a discontinuity point at $k_n = 0$, which is because $W$ jumps from 1 to 0 with the change of $k_n$, the underlying mechanism of one-dimensional chain changes from displacement localization on the left to localization on the right. In the process of $W$ jumping, the transformation of the displacement distribution from the boundary mode to the body mode ($g = 1$) leads to the divergence of $\kappa_u$ at $k_n = 0$.

## 3.2. Geometric linear framework

The excitation method of the start node in section 3.1 is transformed into the force deviating from the center of mass, as shown in Fig. 4(a). In this case, the one-dimensional chain not only has displacement deformation, but also has rotation deformation. First, very small eccentricity is discussed, namely, the angle is very small, and the deformation of the structure can be approximated as a linear small deformation. Therefore, geometric nonlinearity is not considered in FEM, and the results are shown in Fig. 4(a). Under the left and right excitation, the displacement distribution is consistent with Fig. 3(a), with the response increasing exponentially under the left excitation (top figure) and decaying exponentially under the right excitation (lower figure). However, no matter the left or right excitation, the angle fields all appear to decay inward, and the angle distribution remains symmetric when $F_0 = F_N$, presenting reciprocity, as shown in Fig. 4(c).

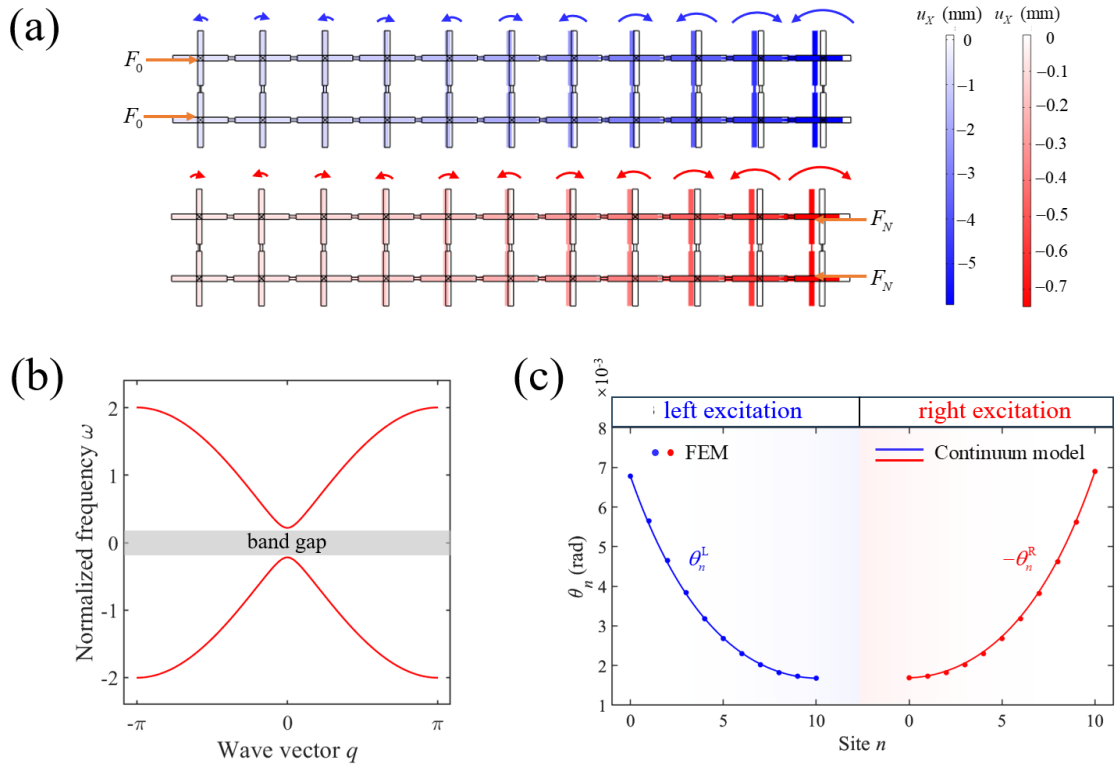

Fig. 4. (a) The top and lower figures are the results of FEM under the left and right eccentric excitation, respectively, where the scaling factors of the top and lower figures are 1 and 10, respectively. The depth of the filling color of the cross represents the magnitude of the displacement field in the X direction, and the arrow and arc length of the arc above the cross represents the direction and magnitude of the angle, respectively. (b) The dispersion relation of the angle. (c) The angle is reciprocal at small deformation, and the blue line (left excitation) and the red line (right excitation) are symmetrical in the figure.

To better analyze and understand the above phenomena, a theoretical model of linear small deformation is established. In the modeling of the geometric linear framework, the angle $\theta_n$ is assumed to be infinitesimal, and the displacement $u_n$ and angle $\theta_n$ are both approximated as a first-order expansion concerning the excitation $F_0$, namely:

$$u_{n+1} - u_n = -(1/k_l) F_{n+1,n}. \tag{6}$$

In Eq. (6), the influence of angle is truncated as higher-order infinitesimal, and Eq.(6) indicates that the distribution of displacement deformation will conform to the distribution of pure displacement in section 3.1.

For angle deformation, the governing equation is expressed as:

$$0 = -k_\theta (\theta_{n+1} + \theta_{n-1} + 4\theta_n) + k_s l^2 (\theta_{n+1} + \theta_{n-1} - 2\theta_n). \tag{7}$$

When the band theory is used to analyze the angle (see Appendix B for details), because the torsion stiffness $k_\theta$ of the shim is non-zero, this results in $\omega(0)$ never being zero, so that the zero frequency is always in the band gap, as shown in Fig. 4(b). This indicates that the angle field at zero frequency corresponds to a no propagating mode, and no matter from which side the excitation is applied, the angle distribution will appear to decay inward. The angular governing equation Eq. (7) under the continuum limit is:

$$\frac{d^2\theta}{dX^2} - (\frac{1}{\lambda_\theta^*})^2 \theta = 0, \tag{8}$$

where $\lambda_\theta^* = \sqrt{(k_s l^2 - k_\theta)/(6k_\theta)}$, it is the characteristic decay length of the angle distribution. Eq. (8) has the spatial symmetry of $X \to -X, \theta \to -\theta$, which confirms that the angle is reciprocal in the case of linear small deformation. The solution of Eq. (8) can be expressed as a linear combination of $\exp(X/\lambda_\theta^*)$ and $\exp(-X/\lambda_\theta^*)$, namely :

$$\theta = c_1 e^{X/\lambda_\theta^*} + c_2 e^{-X/\lambda_\theta^*}. \tag{9}$$

The curves of Eq. (9) are shown in the solid line in Fig. 4(c), where the solution for the left excitation (blue line) and the solution for the right excitation (red line) are symmetrical.

### 3.3. Geometric nonlinear frame

When the eccentric distance of the force is further increased, the one-dimensional chain enters the large deformation of the angle, and the linear small deformation hypothesis in the upper section is no longer applicable. First, the responses of the one-dimensional chain with large deformation under left and right excitation are simulated in the simulation software. At this time, geometric nonlinearity is considered in the simulation, and to ensure that the angle deformation is within a reasonable range, the moment generated by the actuation force should be of the same order of magnitude as $k_\theta$, so $k_n \ll k_l$ is satisfied in the simulation (parameters of simulation in this section, $k_n / k_l = 2 \times 10^{-3}$). The results of FEM show a phenomenon different from section 3.2, namely, the angle field has an asymmetric response, and the comparison between left and right excitation is shown in Fig. 5(a). Next, we focus on the angle distribution under left excitation, and the simulation results show that the characteristics of the angle distribution are dependent on the end displacement $u_N$. When $u_N$ is positive, the angle distributions always decay from the left side (excitation side) to the right side, as shown in Fig. 5(b) and Fig. 5(c), and with the increase of $u_N$, the decay rate is faster, as shown in the inset in Fig. 5(c), the normalized angle distribution curve corresponding to Fig. 5(b) (red line) is always below the normalized curve corresponding to Fig. 5(c) (black line). When $u_N$ is negative, the angle distribution still decays from the left side to the right side when $|u_N|$ is small, as shown in Fig. 5(d); but when $|u_N|$ increases to a certain critical point, the angle distribution changes to increase monotonically from the left side (excitation side) to right side, as shown in Fig. 5(e).

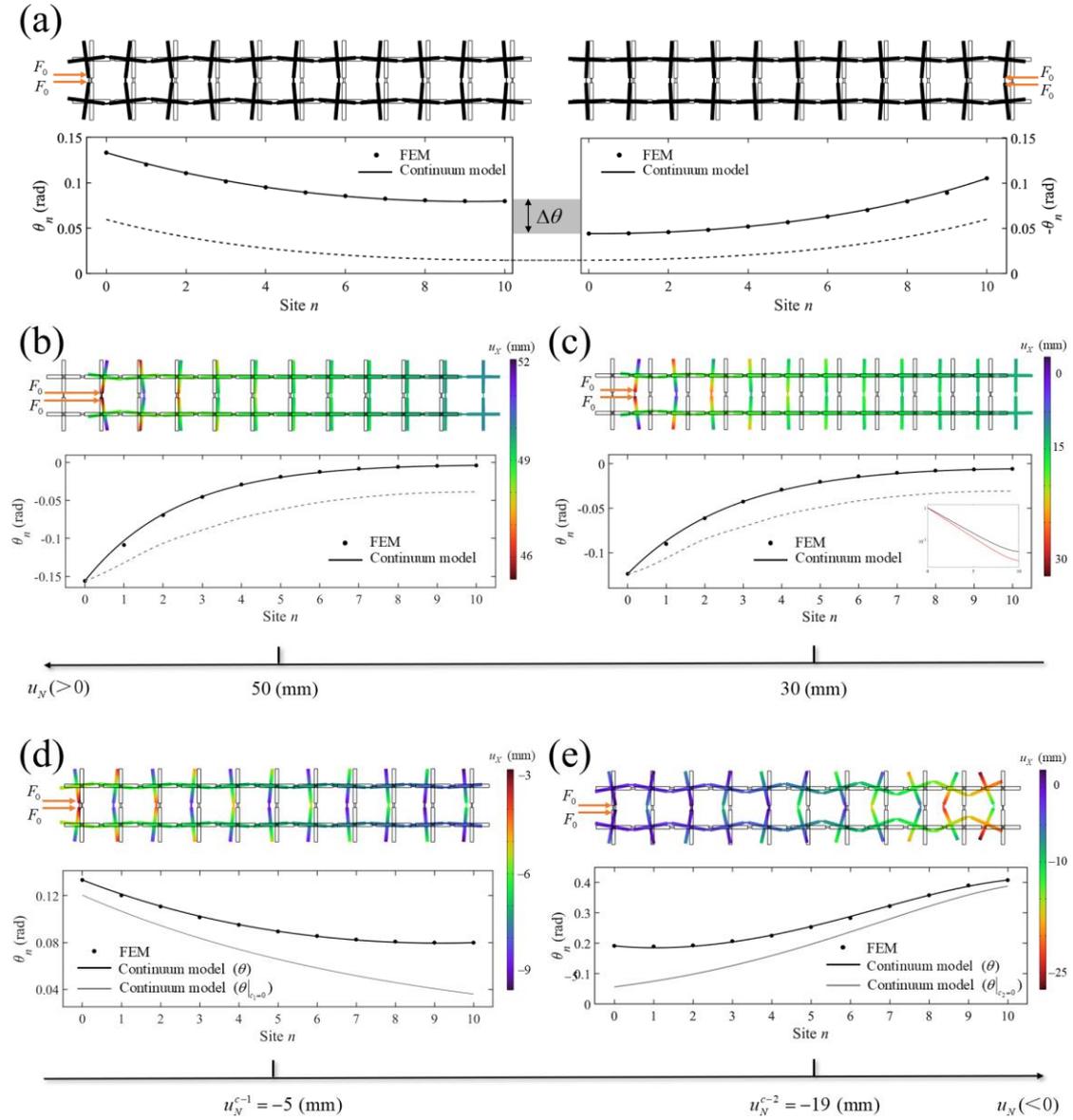

Fig. 5. (a) The top figures show the results of FEM under the left excitation and the right excitation, respectively. The lower figure shows that the responses under the left excitation and the right excitation are asymmetrical, where the black solid dot is the result of FEM and the black solid line is the result of the continuum model (Eq. (15)). The black dashed line is the result of the continuum model (Eq. (9))

without taking into account geometric nonlinearity, where the responses of the left and right excitation are symmetric. (b) and (c) The resulting diagrams corresponding to $u_N > 0$ under the left excitation. (b) and (c) correspond to the cases of $u_N = 50$ (mm) and $u_N = 30$ (mm), respectively, and their angle distribution decays from the left to the right. The distribution curves (black line) are within the gray envelope, where the gray envelope (dashed line) corresponds to the angle distribution of $f_{n,n+1} = 0$. Inset is obtained by normalizing the angle distribution in (b) and (c). (d) and (e) The resulting diagram corresponding to $u_N < 0$ under the left excitation. (d) shows that the angle distribution decays from the left to the right when $u_N = -5$ (mm), and as $u_N$ decreases $-19$ (mm), (e) shows that the angle distribution changes to increase from the left to the right. The solid gray lines are the distribution curves of the dominant deformation mode ($\theta|_{c_1=0}$ or $\theta|_{c_2=0}$).

To investigate the source of the non-reciprocal effect and the cause of decay and increase of two deformation modes, a theoretical model of the geometric nonlinear frame is established. Different from the model of geometric linear frames, in the model of the geometric nonlinear frame, the displacement difference of the centroid of the cross is modeled as:

$$u_{n+1} - u_n = -\frac{F_{n+1,n}}{k_l} - (l + \frac{l_h}{2})(2 - \cos\theta_n - \cos\theta_{n+1}), \tag{10}$$

which is composed of two parts: the tensional deformation of the shim and the change of the horizontal projection generated by the rotation of the cross. According to Eq.(10), the governing equation of displacement is as follows:

$$\mathrm{C}_{(N'-1) \times N'} \mathrm{U} = -k_l l [a_0, ..., a_n, ..., a_{N-1}]^\mathrm{T}, \tag{11}$$

where, $a_n = 2 - \cos\theta_n - \cos\theta_{n+1}, n \in [0, N-1]$. Let $a_N = 0$, the displacement solution can be written as:

$$u_n = c_0 g^{n-N} + (l + \frac{l_h}{2}) \sum_{i=n}^{N} a_i, \tag{12}$$

where $c_0 g^{n-N}$ is the general solution corresponding to the homogeneous form of

Eq.(11) (namely, $CU = 0$), $c_0$ is determined by the boundary conditions; $(l + l_h/2)\sum_{i=n}^{N} a_i$ is the particular solution corresponding to $U(N,1) = 0$, in the process of solving the particular solution, since $k_n \ll k_l$, two elements $c_1$ and $c_2$ in the compatibility matrix C are approximated by $c_1 = c_2 = k_l$, under this particular solution, $c_0 = u_N$. The governing equation for angle is as follows (see Appendix B for details):

$$\frac{d^2\theta}{dX^2} - \frac{(2k_l l u_N g^{-N} \ln g)g^X + 6k_\theta}{k_s l^2 - k_\theta}\theta = 0, \tag{13}$$

where, $g^X$ is generated due to the addition of actuation forces, and it is the addition of this term that makes the symmetry of $X \to -X, \theta \to -\theta$ in Eq.(13) no longer hold, making the angle deformation obtain non-reciprocal effect in nonlinear large deformation, which explains the asymmetric response in Fig. 5(a).

Eq. (13) is further simplified, as $\Delta = 2k_n/(k_l - k_n)$ in $g^X = [1+\Delta]^X$ approaches 0, $g^X$ in Eq. (11) is simplified to $1 + \Delta X$, and a solvable differential equation is obtained:

$$\frac{d^2\theta}{dX^2} - (aX + b)\theta = 0, \tag{14}$$

where,

$$a = (2\Delta k_l l u_N g^{-N} \ln g)/(k_s l^2 - k_\theta),$$
$$b = (2k_l l u_N g^{-N} \ln g + 6k_\theta)/(k_s l^2 - k_\theta).$$

Eq. (14) corresponds to the stationary Schrodinger equation with linear potential energy[39], let $\zeta_s = (aX + b)/|a|^{2/3}$, the solution of Eq. (14) can be expressed as:

$$\theta = c_1 \text{Ai}(\zeta_s) + c_2 \text{Bi}(\zeta_s), \tag{15}$$

Where, $\text{Ai}(\zeta_s)$ and $\text{Bi}(\zeta_s)$ are Airy functions.

Next, two phenomena of decay and increase under left excitation are discussed. In the above modeling, The force exerted by the shim between the *n*-th node and (*n*+1)-th node on the *n*-th cross is (see Appendix B for details) :

$$f_{n,n+1} = k_l u_N g^{-N}(g^{n+1} - g^n). \tag{16}$$

When $u_N > 0$, the force of the internode shim is all tensions. In this case, $f_{n,n+1}$ always exerts a moment on the cross to make the angle tend to 0, which makes the corresponding angle distribution of $u_N > 0$ fall within the envelope formed by the angle distribution of $f_{n,n+1} = 0$, as shown in Fig. 5(b) and Fig. 5(c), so in this case, the angle distribution can only decay from the left side to the right side. At the same time, with the increase of $u_N$, $f_{n,n+1}(u_N)$ also increases, which means an increase of the moment that forces the angle to 0, resulting in a faster decay rate, as shown in the inset in Fig. 5(c).

When $u_N < 0$, the force of the internode shim is all pressure, which meets the necessary condition that the angle increases from the left side to the right side —— the force of the shim between the (N-1)-th node and the N-th node must be pressure (see Appendix B for details). We focus on two critical cases, the first critical is $\theta(N-1) = \theta(N)$, and the $|u_N|$ is small at this time, which is denoting as $u_N^{c-1}$, and $u_N^{c-1}$ is the critical point at which the angle distribution begins to have increasing intervals, as shown in the simulation results in Fig 5(d). The result of Eq.(15) is represented by the black solid line in Fig. 5(d), and the decay deformation mode $\theta|_{c_1=0}$ in Eq. (15) is represented by the gray solid line in Fig. 5(d), the trend of the gray solid line is almost consistent with the black solid line, indicating that the decay mode almost dominates the deformation at this time. The second critical is $\theta(0) = \theta(1)$, at which time $|u_N|$ is larger, and the $|u_N|$ at this time is denoting $u_N^{c-2}$, $u_N^{c-2}$ corresponds to the critical where the angle can increase monotonously from the left side to the right side, as shown in the simulation results in Fig. 5(e). At this time, the increase of deformation mode $\theta|_{c_2=0}$ dominates the angle deformation, as shown in the results of the black solid line and the gray solid line in Fig. 5(e). In summary, with

the decrease of $u_N$, the transition of angle distribution from decay to increase is the result of competition between decay and increase deformation modes in Eq. (15).

In this section, we demonstrate the feasibility of the scheme for transferring non-reciprocity from displacement to angle through coupling under large nonlinear deformation, but more importantly, we find that the amplitude of the angle can also increase with node numbering under appropriate conditions. In the following studies, we try to realize this phenomenon of increasing amplitude in dynamic systems. Inspired by the fact that actuation forces can input energy to the one-dimensional chain in the previous chapter, we attempt to apply this active control force to a dynamic system with constant energy to achieve the phenomenon of increasing angle amplitude, the soliton system supported by the reverse rotating cross/square structure seems to be a good choice.

## 4. Dynamic non-reciprocity

In the study of dynamic behavior, we select a long chain with $N' = 201$ nodes, whose active control force is applied in the same way as in Fig. 5(a), but unlike static loading, in dynamic excitation, a shock excitation is applied to the start section. The governing equations of dynamic are as follows:

$$m\frac{\partial^2 u}{\partial t^2} - k_l \frac{\partial^2 u}{\partial X^2} - k_l(2l+l_h)\theta\frac{\partial \theta}{\partial X} = -2k_n\frac{\partial u}{\partial X}, \quad (17)$$

$$J\frac{\partial^2 \theta}{\partial t^2} + k_\theta(6\theta + \frac{\partial^2 \theta}{\partial X^2}) + 2k_l l[\frac{\partial u}{\partial X} + (l+\frac{l_h}{2})\theta^2]\theta - k_s l^2 \frac{\partial^2 \theta}{\partial X^2} = 0. \quad (18)$$

The left side of the equation system corresponds to the governing equation of vector solitons, while the right side of the equation system has a perturbative term $-2k_n\partial_X u$. The Runge-Kutta method is used in MATLAB to solve the equation system, numerical solutions show that the amplitude of the pulse solitons under the left excitation increases with the increase of the node numbering, as shown in Fig. 6(a). The schematic diagram of the soliton propagation in structure is shown in Fig. 6(c). Then we also carry out shock excitation on the right side, and the spatiotemporal diagram of soliton propagation under left and right excitation correspond to the top and lower figure in Fig. 6(b), respectively. From the spatiotemporal diagram of the right excitation, it can be seen that the amplitude of the soliton is decreasing. Different from the rule that the amplitude of the soliton is continuously increasing with a stable form under the left excitation, the amplitude of the soliton under the right excitation does not change continuously, but suddenly transforms into a strongly dispersive wave when it is below a specific amplitude. In addition, even before the distortion of the solitary wave, the solitary waves corresponding to the left and right excitation are asymmetrical, because the perturbative term breaks the spatial symmetry of the Eqs. (17) and (18).

Soliton systems are systems with constant energy, this means that the energy change of the system will all come from the perturbation term, which can be either actuating like Eq. (17) or dissipative[40, 41]. During the propagation of pulse soliton, the energy changed by the active control force of each node is:

$$\Delta E = 2\int_{-\infty}^{+\infty} -2k_n \frac{\partial u}{\partial X}\frac{\partial u}{\partial T} dT = \begin{cases} 4k_n \int_{-\infty}^{+\infty} (\frac{\partial u}{\partial \zeta_d})^2 d\zeta_d & \text{if } c>0 \\ -4k_n \int_{-\infty}^{+\infty} (\frac{\partial u}{\partial \zeta_d})^2 d\zeta_d & \text{if } c<0 \end{cases}, \quad (19)$$

where, $\zeta_d = X - cT$ is the traveling wave coordinate, and the normalized pulse speed $c$ is positive (negative) under left excitation (right excitation). The result of Eq. (19) shows that the energy of solitons increases (decreases) when $c>0$ ($c<0$). In Appendix C, we prove the positive correlation between the amplitude and energy value, namely, $dA/dE>0$. Therefore, according to the result of Eq. (19), it can be obtained that when solitons propagate from left to right ($c>0$), the active control force injects energy into the chain, thereby increasing the amplitude of the soliton, while the right excitation ($c<0$) is opposite. The soliton solution needs to satisfy a restriction condition that the soliton amplitude should be greater than the amplitude gap $A_p^{\text{gap}}$, the existence of the amplitude gap leads to the sudden distortion when the soliton amplitude is less than $A_p^{\text{gap}}$ under the right excitation. At the same time, due to solitons having a permanent form, even if there are defects in the structure (some units have no active control force), the solitons excited in our active system can still realize unidirectional amplitude amplification with a stable form, as shown in Fig. 6(d).

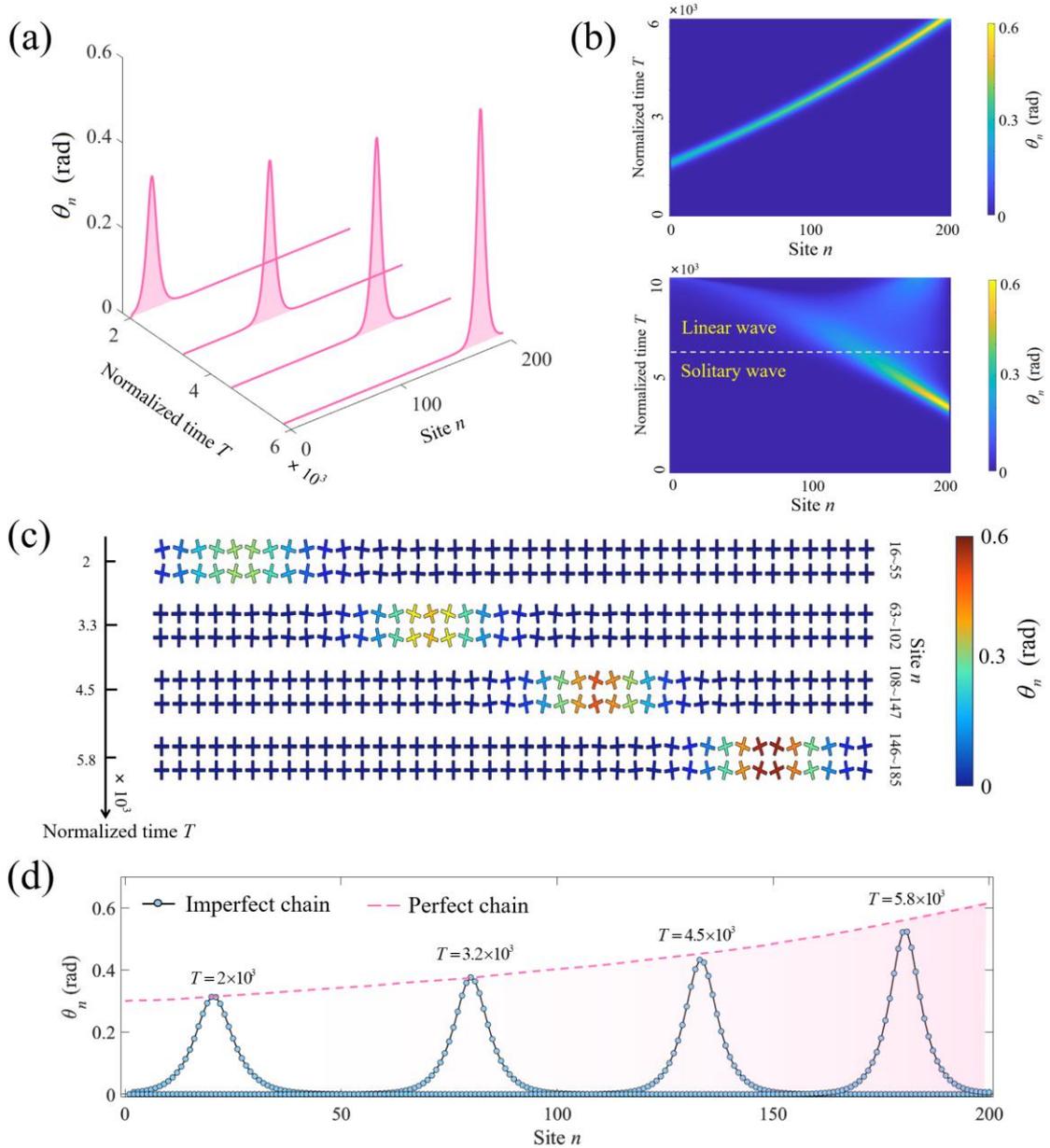

Fig. 6. (a) The amplitude of the soliton increases in the propagation process under the left excitation, and $k_n / k_l = 3.7 \times 10^{-4}$. (b) Solitons maintain a stable form under the left excitation, and the amplitude increases during propagation (top figure). Under the right excitation, the amplitude of soliton decreases during the propagation process and maintains a stable form when the amplitude is greater than the amplitude gap. When the amplitude is reduced to the amplitude gap, the non-dispersive solitary wave transforms into a strongly dispersive linear wave (lower figure). (c) Deformation diagram of numerical solution in structure. (d) A comparison of the skin effects between imperfect structure (no active control force at nodes 100-th to 110-th) and perfect structure.

## 5. Conclusion

To conclude, we propose a robotic metamaterial with two components of rotation and displacement, which has more extreme properties and different combinations of functions. Specifically, in the static behavior, the addition of active control breaks the spatial symmetry of the structure, which induces a topological nontrivial phase, thus achieving unidirectional amplification and asymmetric response in displacement deformation. Then we demonstrate the feasibility of transferring non-reciprocal effects from displacement to angle under nonlinear large deformation, which lays a foundation for the study of dynamic behavior. In the dynamic behavior, amplitude unidirectional amplification and asymmetric transmission are extended to soliton systems. This discovery links the skin effect and soliton science and provides a new platform for the comprehensive regulation of solitary waves. We anticipate that these new characteristics will be expected to be used in crawling robots to overcome the effects of manufacturing errors and dissipation in the operating environment.

**Declaration of Competing Interest**

The authors declare that they have no known competing financial interests or personal relationships that could have appeared to influence the work reported in this paper.

**Acknowledgments**

This work was supported by the National Natural Science Foundation of China (Nos. 12272154, 12172183), the National Key Research and Development Program of China (No. 2023YFE0111000), the Natural Science Foundation of Zhejiang

**Appendix A. Model of the pure displacement framework**

To solve the dispersion relation of displacement under PBC, the form of displacement $u_n$ is set to be a Bloch wave, namely, $u_n = \tilde{u}(q)e^{i(qn-\omega t)}$, and substitute it into Eq.(1):

$$-m\omega^2 = (k_l - k_n)e^{iq} + (k_l + k_n)e^{-iq} - 2k_l. \tag{A-1}$$

The spectrum under PBC is shown in Fig. A. When $k_n \neq 0$, the branches form a loop geometry. For a reference point $\omega$ within the closed loop, the point gap can be well defined, and the corresponding point gap has a spectral winding number characteristic, which is expressed as:

$$W(\omega) = \frac{1}{2\pi i}\oint_C \frac{1}{C(q)-\omega}dC(q), \tag{A-2}$$

where, $C(q) = (k_l - k_n)e^{iq} + (k_l + k_n)e^{-iq} - 2k_l$. For static situations, the reference point is $\omega = 0$, When $k_n = 0$, the loop degenerates into a straight line, where $W(0)=0$, but when $k_n \neq 0$, the integral curve passes through the singularity of the integrand function ($C(0) = 0$), as shown in Fig. A(a) and Fig. A(c), and the integral corresponding to Eq. (A-2) is singular, namely, $W(0)$ cannot characterize the topological properties of $k_n<0$ and $k_n>0$.

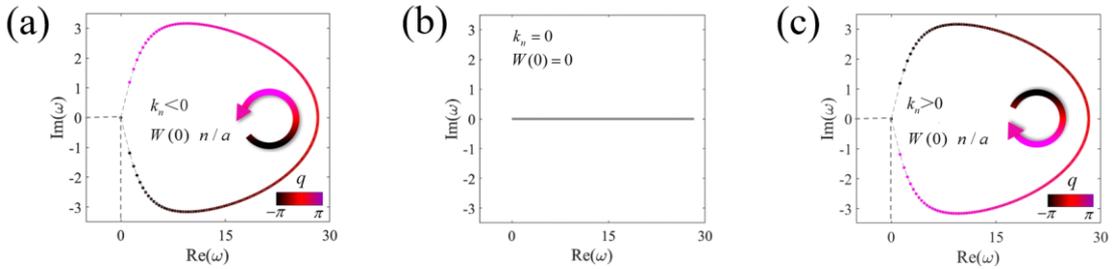

Fig. A Spectra of structures under PBC

OBC can be expressed as applying a rightward force $F_{0,1} = -k_n(u_0 + u_1)$ on the left side, and applying a leftward force $F_{N,N-1} = -k_n(u_{N-1} + u_N)$ on the right side. The modeling diagram of finite structure is shown in Fig. 2(a) in section 3.1. The governing equation of the one-dimensional chain is written as:

$$m\ddot{u}_0 = -(k_l + k_n)u_0 + (k_l - k_n)u_1, \tag{A-3-a}$$

$$m\ddot{u}_n = (k_l - k_n)u_{n+1} + (k_l + k_n)u_{n-1} - 2k_l u_n, n \in [1, N-1], \tag{A-3-b}$$

$$m\ddot{u}_N = (k_l + k_n)u_{N-1} - (k_l - k_n)u_N. \tag{A-3-c}$$

Eq. (A.3) can be written in matrix form:

$$-m\omega^2 U = D_{N' \times N'} U, \tag{A-4}$$

where,

$$D_{N' \times N'} = \begin{bmatrix} -c_1 & c_2 & \cdots & & & 0 \\ 0 & c_2 & -(c_1+c_2) & c_1 & \cdots & 0 \\ 0 & \cdots & c_2 & -(c_1+c_2) & c_1 & \cdots & 0 \\ 0 & \cdots & & c_2 & -(c_1+c_2) & c_1 & 0 \\ 0 & & \cdots & & & c_1 & -c_2 \end{bmatrix}.$$

Eq.(A-4) is numerically solved to obtain the spectrum diagram of the finite structure. In the band gap, an isolated band appears at zero frequency, which is shown in Fig. 2 (b) in section 3.1. According to the analysis in section 3.1, the zero-frequency mode can be reduced to a mode similar to the zero-energy mode, and the governing equation can be reduced to: $F_{n,n+1} = -k_l(u_{n+1} - u_n), n \in [0, N-1]$. Write the above formula in matrix form:

$$C_{(N'-1) \times N'} \begin{bmatrix} u_0 \\ u_1 \\ \vdots \\ u_{N-1} \\ u_N \end{bmatrix} = \begin{bmatrix} c_1 & -c_2 & \cdots & & & 0 \\ 0 & c_1 & -c_2 & \cdots & & 0 \\ 0 & \cdots & c_1 & -c_2 & \cdots & 0 \\ 0 & \cdots & & c_1 & -c_2 & 0 \\ 0 & \cdots & & & c_1 & -c_2 \end{bmatrix} \begin{bmatrix} u_0 \\ u_1 \\ \vdots \\ u_{N-1} \\ u_N \end{bmatrix} = 0, \tag{A-5}$$

where, $C_{(N'-1) \times N'}$ is the compatibility matrix.

In the pure displacement frame, the governing equations of statics are as follows:

$$\text{BC-L: } 0 = k_l(u_1 - u_0) + F_0, \ 0 = (k_l + k_n)u_{N-1} - (k_l - k_n)u_N, \tag{A-6}$$

$$\text{BC-R: } 0 = -(k_l + k_n)u_0 + (k_l - k_n)u, \ 0 = k_l(u_N - u_{N-1}) + F_N, \tag{A-7}$$

$$0 = (k_l - k_n)u_{n+1} + (k_l + k_n)u_{n-1} - 2k_l u_n, \ n \in [1, N-1]. \tag{A-8}$$

Assuming that the envelope of the displacement field has a small gradient, a continuous displacement field $u(X)$ is introduced such that $u_n = u(X_n)$. Performing a Taylor expansion of $u_{n\pm 1} = u(X_n \pm 1)$ at $X_n$:

$$u_{n\pm 1} \approx u(X_n) \pm \left.\frac{\partial u}{\partial X}\right|_{X=X_n} + \frac{1}{2}\left.\frac{\partial^2 u}{\partial X^2}\right|_{X=X_n}. \tag{A-9}$$

Substitute Eq. (A-9) into Eq. (A-8):

$$\frac{d^2 u}{dX^2} - \frac{1}{\lambda_u^*}\frac{du}{dX} = 0, \tag{A-10}$$

where $\lambda_u^* = k_l / (2k_n)$ is the characteristic decay length of the displacement field. Eq. (A-10) is a second-order linear differential equation, whose solution is:

$$u = C_1 e^{X/\lambda_u^*} + C_2 e^{-X/\lambda_u^*}, \tag{A-11}$$

where, $C_1$ and $C_2$ are integral constants determined by the boundary conditions (Eq. (A-6) or Eq. (A-7)). The results of the continuum model are shown in the red solid line in Fig. 2 (d) of section 3.1.

When the left side is excited, the boundary condition BC-L is selected, and the displacement generated by the 0-th node is $u_0^L = -F_0 / [k_l(g-1)]$. According to the geometric distribution law, $u_N^L = -g^N F_0 / [k_l(g-1)]$. When the right side is excited, the boundary condition BC-R is selected, and the 0-th displacement $u_0^R = -g^{-N} F_0 / [k_l(1-g^{-1})]$ is obtained. The output displacement difference between the left excitation and the right excitation is:

$$\Delta u = u_N^L + u_0^R = -\left[\frac{g^N}{k_l(g-1)} + \frac{g^{-N}}{k_l(1-g^{-1})}\right]F_0 = \kappa_u F_0, \tag{A-12}$$

where, $\kappa_u$ is non-reciprocal susceptibility.

**Appendix B. Model of eccentric excitation cases**

(i) Geometrically linear case

The governing equation of angle is:

$$J\frac{\partial^2 \theta_n}{\partial t^2} = -k_\theta(\theta_{n+1} + \theta_{n-1} + 4\theta_n) + k_s l^2(\theta_{n+1} + \theta_{n-1} - 2\theta_n). \tag{B-1}$$

Set $\delta\theta_n$ as the Bloch wave solution, namely, $\theta_n = \tilde{\theta}(q)e^{i(qn-\omega t)}$, and the dispersion relation of the angle is as follows:

$$-J\omega^2 = -k_\theta(e^{iq} + e^{-iq} + 4) + k_s l^2(e^{iq} + e^{-iq} - 2), \tag{B-2}$$

$k_\theta$ is non-zero, which will result in $\omega(q)|_{q=0}$ never being zero, namely, the band gap is always open.

(ii) Geometric nonlinear case

The governing equation from the 1-th node to the (N-1)-th node is:

$$\begin{aligned}0 = &-k_\theta(\theta_{n+1} + \theta_{n-1} + 4\theta_n) \\ &-k_l(f_{n-1,n} + f_{n,n+1})l\sin\theta_n \\ &+ k_s(\sin\theta_{n+1} + \sin\theta_{n-1} - 2\sin\theta_n)l^2\cos\theta_n,\end{aligned} \tag{B-3}$$

where,

$$f_{n,n+1} = k_l[u_{n+1} - u_n + (l + \frac{l_h}{2})(2 - \cos\theta_n - \cos\theta_{n+1})], \tag{B-4}$$

$f_{n,n+1}$ is the tension provided by the shim between the $n$-th node and the $(n+1)$-th node. Substituting Eq. (12) into Eq. (B-4), $f_{n,n+1}$ is reduced to:

$$f_{n,n+1} = k_l u_N g^{-N}(g^{n+1} - g^n), \; n \in [0, N-1] \tag{B-5}$$

Substituting Eq. (B-5) into Eq. (B-3), after a continuous angle field $\theta(X)$ ($\theta_n = \theta(X_n)$) is introduced, the governing equation of the angle is obtained:

$$\frac{d^2\theta}{dX^2} - \frac{(2k_l l u_N g^{-N} \ln g)g^X + 6k_\theta}{k_s l^2 - k_\theta}\theta = 0. \tag{B-6}$$

In the case of left excitation, the governing equation for the $N$-th node can be written as:

$$0 = -k_\theta(\theta_{N-1} + 3\theta_N) + k_s l^2(\sin\theta_{N-1} - \sin\theta_N)$$
$$- k_l[u_N - u_{N-1} + (l + \frac{l_h}{2})(2 - \cos\theta_N - \cos\theta_{N-1})]l\sin\theta_N. \quad \text{(B-7)}$$

According to Eq. (B-5), the tension provided by the shim between the $(N-1)$-th node and the $N$-th node is $f_{N-1,N} = k_l u_N (g - g^{-1})$, and Eq. (B-7) is transformed into:

$$f_{N-1,N} = -\left[\frac{3k_\theta}{l}\frac{\theta(N)}{\sin\theta(N)} + (\frac{k_s l^2 - k_\theta}{l})\frac{(d\theta/dX)|_{X=N}}{\sin\theta(N)}\right]. \quad \text{(B-8)}$$

The angle distribution can only have increasing intervals in two cases, one is that $\theta(N)$ and $(d\theta/dX)|_{X=N}$ are greater than 0 (as shown in Fig. 5(e)), and the other is that $\theta(N)$ and $(d\theta/dX)|_{X=N}$ are less than 0, but the occurrence of these two cases require $f_{N-1,N} < 0$, namely, the force of the shim must be pressure ($u_N < 0$).

**Appendix C. Dynamics model**

The governing equations of the dynamics are as follows:

$$m\frac{\partial^2 u_n}{\partial t^2} = k_n(u_{n-1} - u_{n+1}) + k_l(u_{n+1} + u_{n-1} - 2u_n)$$
$$+ k_l(l + \frac{l_h}{2})(\cos\theta_{n-1} - \cos\theta_{n+1}), \quad \text{(C-1)}$$

$$J\frac{\partial^2 \theta_n}{\partial t^2} = -k_\theta(\theta_{n+1} + \theta_{n-1} + 4\theta_n)$$
$$- k_l(u_{n+1} - u_{n-1})l\sin\theta_n$$
$$- k_l(4 - 2\cos\theta_n - \cos\theta_{n+1} - \cos\theta_{n-1})(l + \frac{l_h}{2})l\sin\theta_n \quad \text{(C-2)}$$
$$+ k_s(\sin\theta_{n+1} + \sin\theta_{n-1} - 2\sin\theta_n)l^2\cos\theta_n,$$

Eqs.(C-1) and (C-2) are normalized as follows: $U = u/(2l)$, $X = x/(2l)$, $T = t\sqrt{k_l/m}$ $K_\theta = k_\theta/k_l l^2$, $K_s = k_s/k_l$, $K_n = k_n/k_l$, $\alpha^2 = ml^2/J$, $\beta = 1 + l_h/(2l)$. The continuum models of Eqs. (C-1) and (C-2) are:

$$\frac{\partial^2 U}{\partial T^2} = \frac{\partial^2 U}{\partial X^2} + \beta\theta\frac{\partial \theta}{\partial X} - 2K_n\frac{\partial U}{\partial X}, \tag{C-3}$$

$$\frac{1}{\alpha^2}\frac{\partial^2 \theta}{\partial T^2} = -K_\theta(6\theta + \frac{\partial^2 \theta}{\partial X^2}) + K_s\frac{\partial^2 \theta}{\partial X^2} - 2[2\frac{\partial U}{\partial X} + \beta\theta^2]\theta. \tag{C-4}$$

In the case of $K_n = 0$, the soliton solution corresponding to Eqs. (C-3) and (C-4) is:

$$\theta = A_p \operatorname{sech}\frac{X - cT}{W_p}, \tag{C-5}$$

$$U = \frac{3K_\theta W_p}{c^2}(1 - \tanh\frac{X - cT}{W_p}), \tag{C-6}$$

where,

$$A_p = \sqrt{\frac{6K_\theta(1 - c^2)}{\beta c^2}}, \tag{C-7}$$

$A_p$ is the amplitude of the pulse soliton,

$$W_p = \sqrt{\frac{K_s - K_\theta - c^2/\alpha^2}{6K_\theta}}, \tag{C-8}$$

$W_p$ is the width of the pulse soliton. The soliton solutions (Eqs. (C-5) and (C-6)) obtained based on the continuum theory, which usually requires that the wavelength is much larger than the length of the unit cell[42], naturally, $W_p>0$ is a necessary condition for the discrete system to support the soliton solutions, which requires $c^2 < \alpha^2(K_s - K_\theta)$. The value range of $c^2$ means that Eq. (C-7) has a minimum value:

$$A_p > \sqrt{\frac{6K_\theta}{\beta}\left[\frac{1}{\alpha^2(K_s - K_\theta)} - 1\right]} = A_p^{\text{gap}}, \tag{C-9}$$

where, $A_p^{\text{gap}}$ is the amplitude gap.

The total energy for the $n$-th pair of crosses can be expressed as:

$$\begin{aligned}e_n &= m(\frac{\partial u_n}{\partial t})^2 + J(\frac{\partial \theta_n}{\partial t})^2 + k_\theta(\theta_n + \theta_{n+1})^2 + \frac{1}{2}k_\theta(2\theta_n)^2 \\ &\quad + k_s l^2(\sin\theta_{n+1} - \sin\theta_n)^2 + k_l[u_{n+1} - u_n + \beta l(2 - \cos\theta_n - \cos\theta_{n+1})]^2,\end{aligned} \tag{C-10}$$

Eqs. (C-5) and (C-6) are substituted into Eq. (C-10) to obtain the energy carried by the soliton:

$$E(c) = \int_{-\infty}^{+\infty} e(X)\, dX = \frac{4}{3} W [k_l l^2 \frac{36 K_\theta^2}{c^2} + 36 k_l l^2 K_\theta^2]$$
$$+ \frac{2}{3} W [k_l l^2 \frac{36 K_\theta^2 (1-c^2)}{\beta \alpha^2 (K_s - K_\theta - c^2/\alpha^2)} + k_s l^2 \frac{36 K_\theta^2 (1-c^2)}{\beta c^2 (K_s - K_\theta - c^2/\alpha^2)}] \quad \text{(C-11)}$$
$$+ 2 W k_\theta \frac{36 K_\theta (1-c^2)}{\beta c^2}.$$

The relationship between $A_p(c)$ and $E(c)$ is shown in Fig. C.

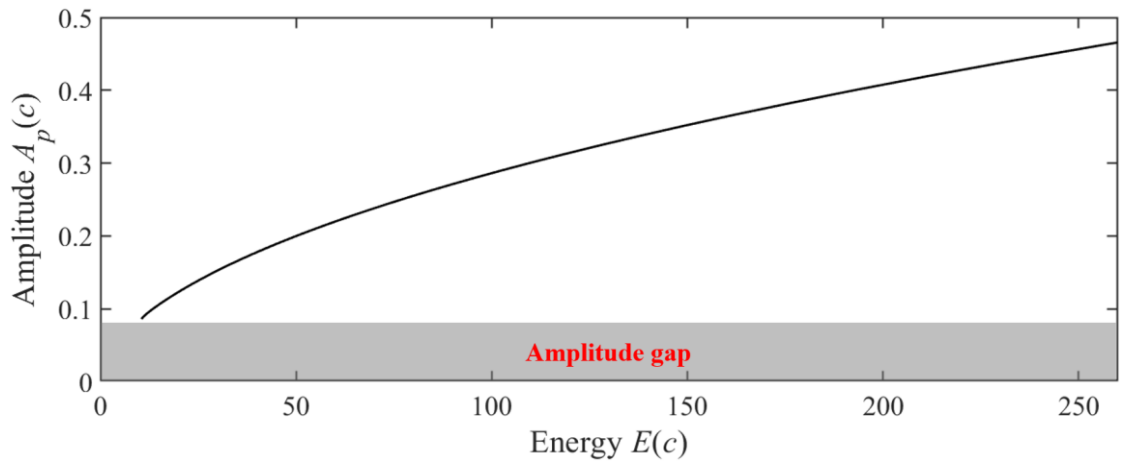

Fig. C  Diagram of the relationship between energy and amplitude